%% file: Main.tex
\documentclass[onecolumn,11pt]{article}

\usepackage[letterpaper,top=0.75in,bottom=0.79in,left=0.75in,right=0.75in]{geometry}
\usepackage[utf8]{inputenc} 
\usepackage[T1]{fontenc}
\usepackage{url}
\usepackage{amsthm,amssymb}
\usepackage{mathrsfs,mathtools,mathabx}
\usepackage{bm}
\usepackage{graphicx}
\usepackage{ifthen}
\usepackage{cite}
\usepackage{amsmath} 
\usepackage{tikz}
\usetikzlibrary{spy}
\usetikzlibrary{arrows,decorations.pathreplacing,patterns}

\newtheorem{theorem}{Theorem}
\newtheorem{remark}{Remark}

\theoremstyle{definition}
\newtheorem{definition}{Definition}

\definecolor{palecgray}{rgb}{0.93, 0.94, 0.94}
\definecolor{chestnut}{rgb}{0.8, 0.36, 0.36}
\definecolor{airforceblue}{rgb}{0.36, 0.54, 0.66}
\definecolor{bazaar}{rgb}{0.6, 0.47, 0.48}
\definecolor{darklavender}{rgb}{0.45, 0.31, 0.59}
\definecolor{bluebell}{rgb}{0.64, 0.64, 0.82}
\definecolor{eggplant}{rgb}{0.38, 0.25, 0.32}
\definecolor{amber}{rgb}{1.0, 0.75, 0.0}
\definecolor{antiquebrass}{rgb}{0.8, 0.58, 0.46}

\begin{document}
% \title{Optimal Function Computation Over Multiple Access Channels Via Hierarchical Constellations} 
\title{Function Computation Over Multiple Access Channels\\ via Hierarchical Constellations}

\author{Saeed Razavikia$^{*}$, Mohammad Kazemi$^{\dagger}$, Deniz Gündüz$^{\dagger}$,  and Carlo Fischione\hspace{-2pt}
\thanks{School of Electrical Engineering and Computer Science, KTH Royal Institute of Technology, Stockholm, Sweden\\
$^\dagger$Department of Electrical and Electronic Engineering, Imperial College London, United Kingdom\\
Email: \{sraz,carlofi\}@kth.se,~\{mohammad.kazemi,d.gunduz\}@imperial.ac.uk} 
    % \thanks{This research was done at the Information Processing and Communications Laboratory, Imperial College London, during the first author’s visit. }
}

\maketitle

\begin{abstract}
We study function computation over a Gaussian multiple-access channel (MAC), where multiple transmitters aim at computing a function of their values at a common receiver. To this end, we propose a novel coded-modulation framework for over-the-air computation (OAC) based on hierarchical constellation design, which supports reliable computation of multiple function outputs using a single channel use. Moreover, we characterize the achievable computation rate and show that the proposed hierarchical constellations can compute $R$ output functions with decoding error probability $\epsilon$ while the gap to the optimal computation rate scales as $\mathcal{O}(\log_2(1/\epsilon)/K)$ for independent source symbols, where $K$ denotes the number of transmitters. Consequently, this gap vanishes as the network size grows, and the optimal rate is asymptotically attained.

Furthermore, we introduce a shielding mechanism based on variable-length block coding that mitigates noise-induced error propagation across constellation levels while preserving the superposition structure of the MAC. We show that the shielding technique improves reliability, yielding a gap that scales optimally as $\mathcal{O}(\log_2\ln{(1/\epsilon}))$ regardless of sources distribution.  Together, these results identify the regimes in which uncoded or lightly coded OAC is information-theoretically optimal, providing a unified framework for low-latency, channel-agnostic function computation.
\end{abstract}

\section{Introduction}

Over-the-air computation (OAC) is a class of joint source--channel coding problems in which multiple transmitters simultaneously access a shared wireless medium. To enable the receiver to recover a function of their messages rather than the individual messages. By exploiting the superposition property of the wireless multiple-access channel (MAC), OAC enables direct function computation at the physical layer, instead of separating communication from computation. Since its introduction in \cite{nazer2007computation}, OAC has been recognized as a key enabler for latency- and bandwidth-efficient operation in large-scale distributed systems, including federated learning~\cite{amiri2020federated}, distributed inference~\cite{yilmaz2025private}, and wireless control~\cite{park2021optimized}.

The OAC literature encompasses both coded and uncoded approaches. Early works focused on linear function computation and were later extended to broader classes of nomographic functions~\cite{goldenbaum2014nomographic}. Analog (uncoded) OAC has attracted particular interest due to its low transceiver complexity and flexibility in supporting different functions. However, conventional analog schemes based on amplitude modulation are highly sensitive to noise and fading. To address these issues, several digital and hybrid schemes have been proposed, including one-bit and few-bit aggregation methods~\cite{zhu2020one}, type-based multiple-access strategies~\cite{qiao2024massive,ngo2024type,tarizzo2025}, and structured designs such as optimization-based in \cite{saeed2023ChannelComp} and codes based on a ring of integers in~\cite{razavikia2024sumcomp}. More recently, constellation-diagram design and optimization methods have been developed to minimize distortion for a given signal-to-noise ratio (SNR) and channel statistics~\cite{razavikia2025towards}. While effective, these approaches are inherently channel-dependent and typically require accurate SNR knowledge at the transceiver.

In this paper, we propose a hierarchical digital constellation scheme that enables the computation of multiple function outputs over noisy MACs using a single channel use. Each transmitter encodes a block of source symbols into a single hierarchical constellation point, which is then transmitted over the channel. The hierarchical structure of the constellation allows the receiver to reliably recover the function values associated with all symbols in the block from the superimposed channel observation. However, this hierarchical construction is inherently susceptible to error propagation across constellation levels. In particular, noise affecting lower-significance levels may propagate to higher-significance levels, thereby corrupting subsequent layers and potentially degrading the recovery of the entire transmitted block.

To address the susceptibility of hierarchical constellations to noise-induced error propagation, we introduce a shielding mechanism inspired by analog coding techniques developed for point-to-point communication~\cite{Chen98Analog,taherzadeh2012single,maddah2024few,maddah2025mystery}. This mechanism protects higher-significance constellation levels from noise contamination originating at lower levels, while preserving the superposition property essential for function computation. The proposed scheme is SNR-agnostic, requiring no knowledge of SNR at either the transmitter or the receiver, and naturally supports high-order modulations.

We characterize the achievable computation rate of the proposed hierarchical OAC scheme and show that it can reliably compute $R$ function outputs with decoding error probability $\epsilon$ while the gap to the optimal computation rate scales as $\mathcal{O}(\log_2(1/\epsilon)/K)$ for independent source symbols at transmitters, where $K$ denotes the number of transmitters. This gap decreases with the network size and vanishes asymptotically, implying that the optimal computation rate is attained in large-scale networks.  Moreover, the shielding mechanism based on variable-length block coding can improve reliability and achieve a near-optimal scaling gap of $\mathcal{O}(\log_2\ln{(1/\epsilon}))$ regardless of the transmitter source distribution.

%****************
\input{Figure/Fig_systemmodel}
%****************

The remainder of the paper is organized as follows. Section~\ref{sec:ProblemForm} introduces the communication model and formulates the problem of function computation over a Gaussian MAC. Section~\ref{sec:Mainresults} presents the proposed hierarchical constellation design and the shielding mechanism that mitigates noise-induced error propagation.  Section~\ref{sec:Theory} provides a performance analysis and characterizes the achievable computation rates of the proposed schemes. Finally, Section~\ref{sec:Conclusion} concludes the paper and discusses future research directions.

\vspace{-5pt}
\section{Problem Formulation}\label{sec:ProblemForm}

We consider a network consisting of $K$ transmitters and a receiver,  where the transmitters communicate with the receiver via a Gaussian MAC.  Transmitter $k$ has a block of symbols $\mathbf{s}_k = [s_k[1],\ldots, s_k[M]]$ with length $M\in \mathbb{Z}_{+}$.  Each symbol $s_k[{m}]$ is a random variable that takes values from a finite set of integers of size $q$, i.e., $s_k[m] \in \mathcal{S}_q:= \{0,1,2,\ldots,q-1\}$ for $k\in [K]$. Symbols $\{s_k{[m]}\}_{m=1}^{M}$ are assumed mutually independent,  with an arbitrary distribution over $\mathcal{S}_q$. The goal of this network is to compute a desired function $f(s_1,\ldots,s_K): \mathcal{S}_q^{K} \mapsto \mathcal{U}$, where $\mathcal{U}$ is a discrete finite set.  The function $f$ is applied symbol-wise across transmitters. More precisely, for each symbol index $m\in [M]$,  the receiver aims to recover $u[m] = f(s_1[m], \ldots,s_K[{m}]) \in \mathcal{U}$.

The computation of function $f$ occurs over a single use of a noisy MAC. Transmitter  $k$ employs  an encoding operator $\mathscr{E}_q(\cdot): \mathcal{S}_q^{M} \mapsto \mathbb{R}$, which maps a block from alphabet $\mathcal{S}_q$ to a channel symbol ${x}_k  \in \mathbb{R}$. We impose an average power constraint on $x_k$, i.e.,   $\mathbb{E}[|{x}_k|^2]\leq 1$, $\forall~k$. Under synchronization assumption among the transmitters, the received signal $y\in \mathbb{R}$  at the receiver is given by
%--------------
 \begin{align}
  \label{eq:received-signal}
     y = \sum\nolimits_{k=1}^{K}x_k + z, 
 \end{align}
%--------------
where $z\sim \mathcal{N}(0,\sigma^2)$ is the additive
white Gaussian noise (AWGN). We employ a decoding operator $\mathscr{D}_{q, K}(\cdot): \mathbb{R} \mapsto \mathcal{U}^{M}$ to compute the output value of the function $f$.  Note that  $\mathscr{D}_{q,K}$ outputs $M$ values, where $\hat{u}[{m}] \in \mathcal{U}$, i.e., 
%--------------
 \begin{align}
    \mathscr{D}_{q,K}(y) = \hat{\mathbf{u}}:= [\hat{u}[1], \hat{u}[2], \ldots, \hat{u}[M]],
 \end{align}
%--------------
where $\hat{u}[{m}]$ is the estimate of  the function evaluated on the $m$-th source symbols, ${u}[{m}]$, $\forall m\in [M]$. 
The system model is depicted in Fig.~\ref{fig:Systemmodel}. We then define the decoding error probability as follows.

\begin{definition}[Decoding Error Probability]
    For a symbol block of length $M$, denoted by $u[1], \ldots , u[M]$,  we define the computation error probability for the first $R$ symbols as
    %----------------
    \begin{align}
        P_{e}(R):= \Pr\Big(\bigcup\nolimits_{m=1}^{R} \hat{u}[m]\neq {u}[{m}]\Big).  
    \end{align}
    %----------------
    The function computation for length $R$ is said to be achievable with error probability $\epsilon >0$ if there exists a code of length $M$ with $P_{e}(R)\leq \epsilon$. 
\end{definition}
\begin{definition}[Computation Rate]\label{def:rate}
  For an SNR-agnostic setting, where neither the transmitters nor the receiver knows the SNR,  the $\epsilon$-computation rate is defined as
   %----------------
 \begin{align}
    \mathcal{R}(\epsilon) :=   \sup_{R\leq  M} \left\{ R \;\middle|\; P_e(R) \le \epsilon \right\},   
\end{align}
%----------------
for a sufficiently large block size of $M$. 
\end{definition}

\subsection{Which Class of Functions?}

Throughout this paper, we focus on the sum function. Specifically, for  all $K$ symbols $s_k[m] \in \mathcal{S}_q$, the desired function is defined as
%------------
\begin{align}
\label{eq:Udefine}
u[m] = f(s_1[m],\ldots,s_K[m]) = \sum\nolimits_{k}s_k[m],
\end{align}
%------------
which implies that the function output alphabet is $\mathcal{U} = \{0,1,\ldots,L-1\}$, where $L:=K(q-1)+1$. 

The proposed framework can be generalized to a wider class of \emph{nomographic} functions of the form $\psi\left(\sum_k \phi_k(s_k)\right)$ by introducing suitable pre-processing and post-processing operations. Specifically, each transmitter applies a pre-processing mapping $\tilde{s}_k=\varphi_k(s_k)\in\tilde{\mathcal{S}}{q_k}= \{a_k + i {c}_k |i= 1,\ldots,q_k\}$ with size $q_k\in \mathbb{Z}_{+}$, for some offset $a_k$, spacing $c_k\neq 0$, and  $\forall  k\in[K]$. The wireless channel then computes the intermediate sum $\tilde{f}=\sum_{k}\varphi_k(s_k)$,  whose output alphabet is given by the Minkowski sum of the input sets, i.e.,  $\tilde{\mathcal{U}}=\bigoplus_{k=1}^K \tilde{\mathcal{S}}_{q_k}=\{\sum_{k}(c_k+a_k),\ldots,\sum_k c_kq_k +\sum_ka_k\}$. Finally, a post-processing function $\psi(\cdot)$ is applied to $\tilde{f}$ to recover the desired nomographic function $f$.

Under this formulation, the constellation size can be adjusted without altering the overall communication procedure.

\begin{remark}
The Kolmogorov–Arnold representation theorem guarantees that any continuous multivariate function admits a representation based on univariate functions combined through summation. Nevertheless, this result is non-constructive and does not provide an explicit or practical procedure to derive a strict nomographic representation of the form $\psi\left(\sum_k \varphi_k(s_k)\right)$ for arbitrary desired functions.
\end{remark}

% \section{Main Results}

\section{The Proposed Hierarchical Constellation}\label{sec:Mainresults}

The proposed scheme is as follows. Transmitter $k$ maps its symbol block $\mathbf{s}_k$  to a real-valued channel codeword $x_k \in \mathbb{R}$ via a shift-map analog encoding scheme~\cite{Chen98Analog}. Specifically, since all source symbols across each block are drawn from the same alphabet $\mathcal{S}_q$\footnote{Note that the source symbols at different transmitters may come from alphabets with different sizes, i.e., $q_1,q_2,\ldots,q_K$ instead of the same $q$. The proposed scheme remains applicable with the only modification being that the base becomes $B:= \sum_{k=1}^K q_k-K+1$.}, each block of symbols can be encapsulated into a single rational (real) number $b_k\in [0,1)$ as follows,
%--------------
\begin{align}
\label{eq:Encoding-real}
    b_k := \sum\nolimits_{m=1}^{M} s_k[{m}] B^{-m}, \quad m\in[M],
\end{align}
%-------------
where  $B$ is a positive integer greater than $L$, i.e., $B\geq L$, to avoid overflow in the superimposed signal during decoding.  Here, each $b_k$ induces a hierarchical pulse amplitude modulation (PAM) constellation with cardinality $q_k^{M}$\footnote{Finite-size hierarchical constellations have been studied in the literature~\cite{vitthaladevuni2003recursive}. However, their reliance on Gray coding leads to destructive constellation overlap in OAC settings~\cite{razavikia2024sumcomp}.}.   Then,  the channel codeword for transmitter $k$ is obtained as $x_k = \eta (b_k - \gamma_k)$, where $\eta$ denotes a power-scaling factor selected to ensure the unit average power constraint on $x_k$. Also,$\gamma_k$ is the average of $b_k$, i.e.,  $\gamma_k = \mathbb{E}[|b_k|]$, $\forall~k\in  [K]$.

\begin{remark}
  Although avoiding overflow only requires the condition $B \geq L$, the base $B$ in \eqref{eq:Encoding-real} serves as a key design parameter governing the noise sensitivity of the scheme~\cite{yoo2017generalized}. Specifically, increasing $B$ enlarges the effective codebook, allowing larger source alphabets $\mathcal{S}_q$. However, this expansion reduces the minimum distance between constellation points, given the power constraint, thereby increasing susceptibility to channel noise.  
\end{remark}

\begin{remark}\label{remakr:progressive}
 Although the base $B$ is assumed to be constant in the proposed encoding scheme, one may alternatively adopt a progressive-base construction as in~\cite{maddah2024few}, where the base varies with the digit index according to $B^{-m} \rightarrow (B-1)!/(B + m- 1)!,$  $\forall~m \geq 1$. Under this formulation, the shift-map representation in \eqref{eq:Encoding-real} is modified to $b_k = \sum_{m=1}^{M} s_k[{m}]/ (B_m)!$, which enables variable alphabet sizes across the symbol index $m$. Moreover, for short-length block, it has been empirically demonstrated that generalized shift-map codes constructed using a set of pairwise coprime integers can further reduce distortion~\cite{yoo2017generalized}.
\end{remark}

%****************
\input{Figure/Fig_Coding}
%****************

\subsection{Extracting Symbols}

 From the received signal in \eqref{eq:received-signal}, the receiver computes $d = \bar{\gamma}+ \frac{y}{\eta}$  and evaluates the expansion in base $B$, i.e., 
%------------
\begin{align}
  d& = \bar{\gamma}+ \frac{y}{\eta}= \sum\nolimits_{k=1}^{K}b_k + \frac{z}{\eta}=: \sum\nolimits_{m=1}^{M} d[m] B^{-m},\label{eq:d-deocded}
\end{align}
%------------
 where $\bar{\gamma} := \sum_{k}\gamma_k$.  The coefficients ${d[m]}$ are obtained via a long-division procedure. Based on these digits, the estimate of $u[m]$ defined in \eqref{eq:Udefine} is recovered as 
%----------------
\begin{align}
    \hat{u}[m] = d[m], \quad m\in [M].
\end{align}
%----------------
Notably, no distributional assumption is imposed on the digits ${u[m]}$. If such statistical information is available, one may further enhance the decoding performance by adopting maximum a posteriori estimation, since the superposition of random variables generally deviates from a uniform distribution.

% The overall coding procedure is depicted in Fig~\ref{fig:Systemmodel}. 

\begin{remark}
The channel model is assumed to be real-valued, and the encoding and decoding operations in \eqref{eq:Encoding-real} and \eqref{eq:d-deocded}, respectively, are performed over the real domain. Nevertheless, the proposed framework can be extended to the complex domain. In that case, hierarchical QAM constellations are used instead of PAM. This extension effectively doubles the computation rate per channel use, at the expense of increased distortion.
\end{remark}

\subsection{Shielded Transmission}\label{sec:Alanog-codes}

So far, the proposed scheme remains susceptible to error propagation, in which channel noise affecting a single symbol in the block \(u\) may spread to subsequent symbols and contaminate the entire block~\cite{maddah2025mystery}. In particular, noise affecting \(u[m],u[m+1],\ldots\) can lead to erroneous decoding of previously recovered symbols \(u[m-1],u[m-2],\ldots,u[1]\). 
To suppress such error propagation, inspired by the analog codes in point-to-point communications~\cite{maddah2025mystery,maddah2024few,klein2016low}, we present a general way of shielding for computing the sum function. More precisely, rather than directly transmitting the scaled version of \(s_k[m]\), each symbol is encoded as
%-----------
\begin{align}
\label{eq:encoded-schem}
    t_k[m] = \frac{\beta_mK^{-1} +  s_k[m]}{(B+\alpha_m)^{m}},
\end{align}
%-----------
where \((\beta_m,\alpha_m)\in\mathbb{Z}_{+}^3\) are design parameters. The parameter \(\alpha_m \geq 0\) enlarges the effective codebook, while \(\beta_m\) introduces a guard offset that restricts noise leakage from less significant to more significant digits.  To prevent aliasing across digits, \(\beta_m\) must satisfy \(0 \leq \beta_m \leq \alpha_m + 1\).
For example, choosing \((\beta_m,\alpha_m)=(1,2)\) yields
%-----------
\begin{align}
   \label{eq:t-coded} 
    t_k[m]=\frac{s_k[m]+1/K}{(B+2)^m}, \quad \forall~m,
\end{align}
%-----------
where the numerator of the sum $\sum_{k=1}^{K}t_k[m]$ is contained in \(\{1,\ldots,B\}\), leaving boundary levels unused within the enlarged alphabet \(\{0,1,\ldots,B+1\}\).  An illustration for the coding scheme in \eqref{eq:t-coded} is depicted in Fig.~\ref{fig:Coding}.  Next, each transmitter  transmits the entire encoded block as
%------------
\begin{align}
    b_k = \sum\nolimits_{m=1}^{M} t_k[m], \quad m\in [M].
\end{align}
%------------
As in the uncoded case, the channel codeword is given by \(x_k=\eta(b_k-{\gamma}_k)\). At the receiver, after compensating for the known shift and scaling, the superimposed signal admits the expansion
%------------
\begin{align}
   \nonumber
   \mathscr{D}_{q, K}(y) 
   :=\hspace{-3pt} \sum\nolimits_{m=1}^{M}\frac{r[m]}{(B+\alpha_m)^m}.
\end{align}
%------------
Given the recovered digits \(r[m]\), the estimate of the desired output symbol is obtained as
%------------
\begin{align}
    \hat{u}[m] = {r[m]-\beta_{m}},\quad \forall\, m\in [M],
\end{align}
%------------
 The encoding rule in~\eqref{eq:encoded-schem} is sufficiently general to encompass both fixed-length and variable-length block coding strategies proposed in~\cite{maddah2025mystery,maddah2024few}. In particular, by employing a variable base \(B^{-m} \rightarrow  (B+m-1)^{-m}\) and choosing \((\beta_m,\alpha_m)=(1,2)\), the progressive coding scheme of~\cite{maddah2024few} is recovered.

\begin{remark}
Beyond limiting error propagation, the offset parameter $\beta_m$ in the encoding rule in \eqref{eq:t-coded} enables explicit error detection and localization across significance levels. 
Specifically, the presence of unused guard levels induced by $\beta_m$ implies that any carry-over caused by channel noise may lead to an invalid digit pattern associated with the guard levels in the decoded coefficient $r[m]$. 
By sequentially checking the consistency of $r[m]$ with the admissible range determined by $\beta_m$, the receiver can identify the smallest index $m$ at which a violation occurs.
\end{remark}

\section{Achievable Computation Rate}\label{sec:Theory}

We analyze the reliability of function computation in terms of the bit error rate defined in Definition~\ref{def:rate}. Our objective is to characterize the maximum number of reliably decoded symbols (digits) under a target error probability. The analysis largely follows the approach developed in~\cite{maddah2025mystery}.

\subsection{Unshielded Transmission}

We first consider the setting, where no guard intervals are employed for the source symbols, i.e., $(\beta_m,\alpha_m)=(0,0),$  $\forall m$. In this case, we have the following result.

\begin{theorem}\label{thm:special}
Consider the hierarchical OAC scheme for a network of size $K$, where each transmitter takes values from a finite set of integers  $\mathcal{S}_q$.  Then,  the $\epsilon$-computation rate admits the form
\begin{align}
\mathcal{R}_1(\epsilon) \leq \frac{\tfrac{1}{2}\log_{2}{\mathrm{SNR}}}{\log_2{q} + \log_2{K}}
-\frac{\log_{2}\!\Big(\frac{c_0Q(1)}{\epsilon}\Big)}{K},
\end{align}
where $c_0>0$ is a positive constant.    
\end{theorem}

\begin{proof}
    
Consider the proposed transmission of a block of $M$ symbols, and suppose that the first $R\leq M$ symbols must be decoded correctly with probability at least $1-\epsilon$. After normalization and offset compensation, the receiver obtains $ y/\eta+\bar{\gamma}=\tilde{x}+\tilde{z}$,  where $\tilde{x}=\sum_{m=1}^{M}u[m]B^{-m}$ and  $\tilde{z}=\sum_{m=1}^{\infty}z[m]B^{-m}$. Here,  $z[m], u[m]\in\{0,1,\ldots,B-1\}$, and $\tilde{z}\sim\mathcal{N}(0,1/\mathrm{SNR})$ where $\mathrm{SNR}=\frac{\eta^2}{\sigma^2}$.  Define $p_m$ as the probability that \emph{the first nonzero} base-$B$ digit of $\tilde{z}$ occurs at position $m$, i.e., 
 $ {p}_m:=\Pr(z[m]\neq 0, z[i]=0, \forall~i <m)$, which is given by
 %--------------
 \begin{align}
  \nonumber
     p_m &= \Pr(B^{-m}\leq |\tilde{z}|< B^{-m+1}),  
     \\
     &=2Q\!\left(\frac{\sqrt{\mathrm{SNR}}}{B^{m}}\right)- 2Q\!\left(\frac{\sqrt{\mathrm{SNR}}}{B^{m-1}}\right), ~~\forall\,m. 
 \end{align}
 %--------------
The probability of error in decoding the first $R$ symbols is decomposed as
 \begin{align}
 \label{eq:pe_r}
    P_e(R) =   P_e\Big(R~\big\vert ~|\tilde{z}|\geq \frac{1}{B^{R}}\Big) +  P_e\Big(R~\big\vert ~|\tilde{z}|< \frac{1}{B^{R}}\Big). 
 \end{align}
The first term corresponds to the event that noise directly corrupts at least one of the first $R$ digits, yielding
 %--------------
\begin{align}
    P_e\Big(R~\big\vert ~|\tilde{z}|\geq \frac{1}{B^{R}}\Big)= 2Q\!\left(\frac{\sqrt{\mathrm{SNR}}}{B^{R}}\right).  \label{eq:pe_r1}  
\end{align}
 %--------------
 When $|\tilde{z}|< B^{-R}$, direct digit errors do not occur. However, an error may still arise if a carry or borrow propagates from a lower-significance digit $m>R$ to digit $R$. Such propagation requires that all intermediate digits take extreme values in their base $B$, namely $B-1$ when $\tilde{z}>0,$ or $0$ when $\tilde{z}<0$ \cite{maddah2025mystery}. For any symmetric distribution over $\mathcal{U}$, we define $\tilde{p}$ below
%----------------
\begin{align}
\tilde{p}^{-1}:=\Pr(u[m]=0)= \Pr(u[m]=B-1), \forall\,m, 
\end{align}
%----------------
which is independent of $m$. The probability that an error originating at digit $m$ propagates to digit $R$ is then  ${1}/{\tilde{p}^{m-R}}$.
Consequently, we obtain
 %--------------
\begin{align}
    \nonumber
    P_e\Big(R\big\vert |\tilde{z}|< \frac{1}{B^{R}}&\Big)= \hspace{-5pt}\sum_{m=R+1}^{M}   \frac{{p}_m}{\tilde{p}^{m-R}} + \hspace{-5pt}\sum_{m=M+1}^{\infty}  \frac{{p}_m}{\tilde{p}^{M-R}}, \\
    & = c_0\hspace{-7pt}\sum_{m=R+1}^{M+1} \hspace{-3pt}  \frac{Q(\frac{\sqrt{\mathrm{SNR}}}{B^{m}})}{\tilde{p}^{m-R}} - \frac{2Q(\frac{\sqrt{\mathrm{SNR}}}{B^{R}})}{\tilde{p}}, \label{eq:pe_r2}
\end{align}
 %--------------
where $c_0:= 2(1-\tilde{p}^{-1})>0$ and the last equality is due to the telescoping sum.  
Next, by substituting  \eqref{eq:pe_r1} and \eqref{eq:pe_r2} into \eqref{eq:pe_r}, the overall error becomes
%----------------
\begin{align}
    P_e(R) \hspace{-1pt}  =  \hspace{-1pt}   2Q\!\Big(\frac{\sqrt{\mathrm{SNR}}}{B^{R}}\Big) \hspace{-2pt}  + \hspace{-1pt}  c_0\hspace{-5pt}\sum_{m=R+1}^{M+1} \hspace{-3pt}  \frac{Q(\frac{\sqrt{\mathrm{SNR}}}{B^{m}})}{\tilde{p}^{m-R}}
     \hspace{-2pt}   - \hspace{-2pt}  \frac{2Q(\frac{\sqrt{\mathrm{SNR}}}{B^{R}})}{\tilde{p}}   
     = c_0\sum\nolimits_{m=R}^{M+1} \frac{Q(\frac{\sqrt{\mathrm{SNR}}}{B^{m}})}{\tilde{p}^{R-m}}, \label{eq:upper_error}
\end{align}
%----------------
where the last inequality comes from merging the first and last terms into the series.  

Next, let $m^{\star}:=\lfloor \log_{B}\sqrt{\mathrm{SNR}}\rfloor$, so that $\sqrt{\text{SNR}}/B^{m^{\star}}\approx 1$. Then, we have 
%----------------
\begin{align}
P_e(R)\;\geq  c_0 \frac{Q(\frac{\sqrt{\mathrm{SNR}}}{B^{m^{\star}}})}{ \tilde{p}^{R-m^{\star}}}=  {c_0}Q(1){\tilde{p}}^{m^\star-R}.
\end{align}
%----------------
 Imposing $P_e(R)\le\epsilon$  yields 
%----------------
\begin{align}
R \;\le\; {m^\star}-\log_{\tilde{p}}\!\Big(\frac{c_0Q(1)}{\epsilon}\Big).
\end{align}
%----------------
Using Definition~\ref{def:rate}, the $\epsilon$-computation rate therefore satisfies 
%---------------
\begin{align}
\tilde{\mathcal{R}}_1(\epsilon) \leq  \frac{\log_{2}{\mathrm{SNR}}}{2\log_2{B}}
-\frac{\log_{2}\!\Big(\frac{c_0Q(1)}{\epsilon}\Big)}{\log_2{\tilde{p}}}. 
\end{align}
%--------------
Finally, for $B\approx Kq$ and independent source symbols,  $\tilde{p}={{q}}^{K}\leq 2^{K}$,\footnote{The upper bound holds without requiring symmetry. Because the analysis yields a lower bound on $P_e(R)$, $\tilde p^{-1}$ can be conservatively replaced by $\min\{\Pr(u[m]=0),\Pr(u[m]=B-1)\}$, which is always upper bounded by $1/2$.} which completes the proof.
\end{proof}

Note that the resulting rate $\mathcal{R}_1(\epsilon)$   coincides with the computation rate reported in \cite{goldenbaum2014nomographic}, up to an additive gap term. Unlike the asymptotic regime considered in \cite{goldenbaum2014nomographic}, this gap arises from the single-shot nature of the proposed scheme.

 The gap term in Theorem~\ref{thm:special} is $\mathcal{O}(K^{-1})$, which decreases as the number of transmitters $K$ grows, reflecting the diminishing probability of carryover/error propagation in large networks.  As a result, in a large network with independent source symbols, unshielded transmission becomes asymptotically optimal. 
 %While enlarging the source alphabet size  $q$ further reduces this gap, it simultaneously degrades the achievable rate, leading to an inherent trade-off. 

\begin{remark}
As $K\to\infty$, the superimposed alphabet over the MAC converges to a Gaussian distribution by the central limit theorem. In this regime, uncoded transmission is known to be optimal for Gaussian sources~\cite{gastpar2008uncoded}.
\end{remark}

\subsection{Shielded Transmission}

We next consider a block-coded design with fixed guard intervals to prevent error propagation regardless of transmitter source distribution.   To preserve symmetry with respect to channel noise, we choose
\begin{align}
\label{eq:beta-bar}
(\beta_1,\alpha_1) = (0,0), \quad  
(\beta_m,\alpha_m) = (\bar{\beta},2\bar{\beta}), \quad m\ge 2,
\end{align}
where $\bar{\beta} \geq 1$. In this setting, digit errors do not propagate across the block. 

\begin{theorem}\label{thm:general_rate}
Consider the hierarchical OAC scheme with base $B$ and encoding parameters $(\beta_m,\alpha_m)$ from \eqref{eq:beta-bar}. Then,  the $\epsilon$-computation rate admits the form
\begin{align}
\mathcal{R}_2(\epsilon)
\geq
\frac{\tfrac{1}{2}\log_2(\mathrm{SNR})}{\log_2(B+2\bar{\beta})}
-
\mathcal{G}_2(\epsilon),
\end{align}
where $\mathcal{G}_2(\epsilon)$ is given by 
\begin{align}
    \mathcal{G}(\epsilon):= \frac{1 }{2\log_2(B+2\bar{\beta})}
\log_2\!\Big(2\ln\!\big(\frac{1}{\epsilon}\big)\Big).
\end{align}
\end{theorem}

\begin{proof}
Since error propagation is eliminated in this setting, the correct decoding of the first $R$ symbols (digits) is violated only when the effective noise exceeds the resolution of the $R$-th digit. Accordingly, the decoding error probability is given by
%---------------
\begin{align} 
\nonumber P_e(R)=
\Pr\!\left(|z| \ge \frac{1}{(B+2\bar{\beta})^R}\right) 
=2Q\!\left(\frac{\sqrt{\mathrm{SNR}}}{(B+2\bar{\beta})^R}\right).
\end{align}
%---------------
Using the Chernoff bound $Q(x)\le \tfrac{1}{2}e^{-x^2/2}$, we obtain
%---------------
\begin{align}
P_e(R)
\le
\exp\!\left(-\frac{\mathrm{SNR}}{2(B+2\bar{\beta})^{2R}}\right)
\le \epsilon,
\end{align}
%---------------
which is equivalent to
\begin{align}
(B+2\bar{\beta})^{2R}\ge \frac{\mathrm{SNR}}{2\ln\!\left(\tfrac{1}{\epsilon}\right)}.
\end{align}
%---------------
Taking logarithms in base $2$ on both sides gives
%---------------
\begin{align}
R \geq \frac{\log_2(\mathrm{SNR})-\log_2(2\ln{1/\epsilon})}{2\log_2(B+2\bar{\beta})},  
\end{align}
%---------------
which completes the proof.

\end{proof}

Theorem \ref{thm:general_rate} shows that introducing guards improves the scaling of the gap term $\mathcal{G}_2(\epsilon) = \mathcal{O}(\log_2\ln{(1/\epsilon)})$  in terms of $\epsilon$ by the order of magnitude. However, it reduces the effective rate due to constellation expansion by a factor of ${\log_2 B}/{\log_2(B+2\bar{\beta})}$.

 To mitigate the tradeoff stated in Theorem \ref{thm:general_rate} with parameter $\bar{\beta}$, variable-length block coding (variable size $B_m$ across the symbol block) can be employed, following~\cite{maddah2025mystery}. Specifically, guards are inserted at positions
%---------------
\begin{align}
m_j = \tfrac{j(j+3)}{2}, \quad B^{-m} \rightarrow \frac{(B-1)!}{(B+m-1)!},
\end{align}
%---------------
so that the number of guard symbols among the first $\mu$ positions grows as $O(\sqrt{\mu})$. Consequently, the number of information digits satisfies $R=\mu-O(\sqrt{\mu})$.%---#############
Specifically, let $J(\mu)$ be the number of guards among the first $\mu$ positions. Since guard indices are located at $m_j$,  the condition $\mu \geq m_j$ implies
%---------------
\begin{align}
    \frac{j(j+3)}{2}\leq \mu~\rightarrow~j^2+3j-2\mu\leq 0~\rightarrow~j\leq \frac{-3+\sqrt{9+8\mu}}{2}. 
\end{align}
%---------------
Consequently, the number of guard digits satisfies $J(\mu)\leq \frac{-3+\sqrt{9+8\mu}}{2} \leq \sqrt{2\mu}$. Therefore, the number of information digits among the first $\mu$ positions is
%---------------
\begin{align}
\label{eq:rate-mu}
    R = \mu -J(\mu) \geq \mu-\sqrt{2\mu}. 
\end{align}
%---------------
Next, note that decoding of the first $\mu$ digits fails only if the effective noise magnitude exceeds the resolution associated with the $\mu$-th digit. Therefore, the decoding error probability admits the bound
\begin{align}
  \nonumber P_e(R)=
\Pr\!\left(|z| \ge \frac{1}{B^{\mu}}\right) 
=2Q\!\left(\frac{\sqrt{\mathrm{SNR}}}{B^{\mu}}\right) \leq   \exp\!\left(-\frac{\mathrm{SNR}}{2B^{2\mu}}\right)
\le \epsilon,
\end{align}
%---------------
Which yields
\begin{align}
    \mu \geq \frac{\frac{1}{2}\log_2{(\mathrm{SNR})}}{\log_2B}-   \frac{\tfrac{1}{2} \log_2(2\ln{(1/\epsilon)})}{\log_2B}. \label{eq:lowerbound-mu}
\end{align}
%---------------
Finally, combining \eqref{eq:rate-mu} and \eqref{eq:lowerbound-mu} establishes the following lower bound on the achievable computation rate.
%---#############
%Repeating the block-coded analysis with $R$ replaced by $\mu$ yields the achievable rate
%---------------
\begin{align*}
\mathcal{R}_3(\epsilon)
\gtrsim
\frac{\tfrac{1}{2}\log_2(\mathrm{SNR})}{\log_2{B}} - \mathcal{G}_3(\epsilon) - o(\log \mathrm{SNR}), 
\end{align*}
%---------------
where 
\begin{align}
    {\mathcal{G}_3(\epsilon):=} \frac{\tfrac{1}{2}\log_2\!\Big(2\ln(\tfrac{1}{\epsilon})\Big)}{\log_2{B}}.
\end{align}
Thus, $\mathcal{R}_3(\epsilon)$ achieves a near-optimal scaling in terms of~$\epsilon$.

\section{Conclusion}\label{sec:Conclusion}
This paper examined function computation over Gaussian MACs. We introduced a digital OAC framework based on hierarchical constellations. We showed that, under independent source symbols, the proposed scheme achieves the optimal computation rate when the network size is large, even with a single channel use. For small-scale networks or arbitrary source distributions, we developed a generalized shielded version of the hierarchical constellation that mitigates noise-induced error propagation while maintaining high spectral efficiency.

Future work will consider extensions to fading channels in which neither transmitters nor receiver possess channel state information. In this context, polynomial-based signaling techniques, such as Huffman polynomial constructions~\cite{walk2019mocz}, represent a promising direction toward improving robustness and further enhancing the realism of the system model.

\bibliographystyle{IEEEtran}
\bibliography{IEEEabrv,Ref}

\end{document}

%% file: Figure/Fig_systemmodel.tex
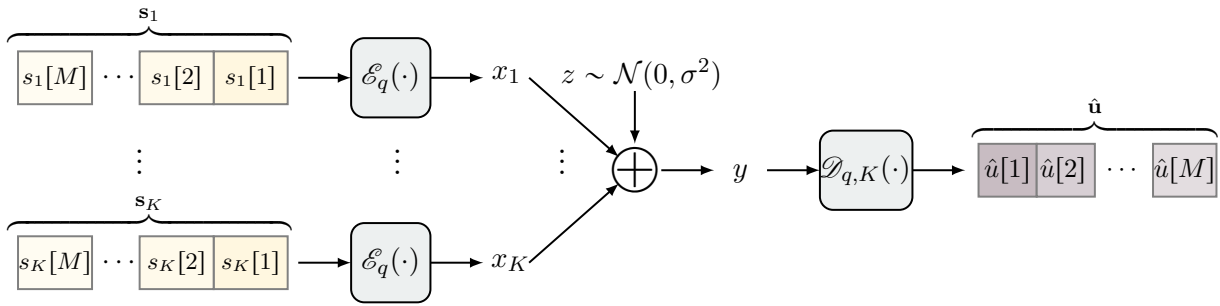
\begin{figure*}[!t]
    \centering

\scalebox{1}{
\tikzset{every picture/.style={line width=0.75pt}} 
\begin{tikzpicture}[x=0.75pt,y=0.75pt,yscale=-1,xscale=1]

% Rect- D(\cdot)
\draw[fill=palecgray , rounded corners=5pt] (200pt, 95pt) rectangle (170pt, 65pt) {};
\draw (185pt,80pt) node   {$\mathscr{E}_{q}(\cdot)$};

\draw [-latex]    (150pt,80pt) -- (170pt,80pt) ;

%--------------
\draw[-latex]    (200pt,150pt) -- (220pt,150pt) ;

\draw[fill=palecgray , rounded corners=5pt] (200pt, 165pt) rectangle (170pt, 135pt) {};
\draw (185pt,150pt) node   {$\mathscr{E}_{q}(\cdot)$};

\draw [-latex]    (150pt,150pt) -- (170pt,150pt) ;

% \draw (210,100) node {$s_1$};
\draw (305,105) node { $x_1$};

\draw (125,80) node {$\overbrace{~~~~~~~~~~~~~~~~~~~~~\quad \qquad }^{\mathbf{s}_1}$};

\draw[color=gray, fill=amber!12 ] (146pt, 90pt) rectangle (118pt, 70pt) {};
\draw[color=gray, fill=amber!8 ] (118pt, 90pt) rectangle (90pt, 70pt) {};
\draw[color=gray, fill=amber!5 ] (72pt, 90pt) rectangle (44pt, 70pt) {};

\draw (132pt,80pt) node {\footnotesize $s_1{[1]}$};
\draw (104pt,80pt) node {\footnotesize  $s_1{[2]}$};
\draw (83pt,80pt) node {  $\cdots$};
\draw (58pt,80pt) node {\footnotesize  $s_1{[M]}$};

\draw[color=gray, fill=amber!12 ] (146pt, 160pt) rectangle (118pt, 140pt) {};
\draw[color=gray, fill=amber!8 ] (118pt, 160pt) rectangle (90pt, 140pt) {};
\draw[color=gray, fill=amber!5 ] (72pt, 160pt) rectangle (44pt, 140pt) {};

\draw (132pt,150pt) node {\footnotesize $s_K{[1]}$};
\draw (104pt,150pt) node {\footnotesize  $s_K{[2]}$};
\draw (83pt,150pt) node {  $\cdots$};
\draw (58pt,150pt) node {\footnotesize  $s_K{[M]}$};

\draw (120,145) node {\Large $\vdots$};
\draw (250,145) node {\Large $\vdots$};

\draw (125,175) node {$\overbrace{~~~~~~~~~~~~~~~~~~~~~\quad \qquad }^{\mathbf{s}_K}$};
\draw (307,200) node { $x_K$};
\draw [-latex]  (200pt,80pt) -- (220pt,80pt) ;

\begin{scope}[shift={(-1.5cm,0)}]
    \draw [-latex]    (330pt,115pt) -- (350pt,115pt) ;

\draw (320pt, 115pt) node {\LARGE $\bigoplus$};

\draw (390,145) node {\Large $\vdots$};
\draw (430,105) node {$z\sim \mathcal{N}(0,\sigma^2)$};

\draw [-latex]  (320pt,85pt) -- (320pt,105pt) ;

%--------------
\draw [-latex]  (280pt,80pt) -- (314pt,110pt) ;

%--------------
\draw [-latex]  (280pt,150pt) -- (314pt,120pt) ;

\draw (480,155) node {$y$};

\draw [-latex]    (370pt,115pt) -- (390pt,115pt) ;

\draw (494pt,95pt) node {$\overbrace{~~~~~~~~~~~~~~~~~\quad \qquad }^{\hat{\mathbf{u}}}$};

\draw[color=gray, fill=eggplant!35 ] (472pt, 125pt) rectangle (450pt, 105pt) {};
\draw[color=gray, fill=eggplant!25 ] (494pt, 125pt) rectangle (472pt, 105pt) {};
\draw[color=gray, fill=eggplant!18 ] (540pt, 125pt) rectangle (516pt, 105pt) {};

\draw (461pt,115pt) node {\small $\hat{u}{[1]}$};
\draw (482pt,115pt) node {\small  $\hat{u}{[2]}$};
\draw (528pt,115pt) node {\small  $\hat{u}{[M]}$};
\draw (505pt,115pt) node {\small  $\cdots$};

\draw[fill=palecgray , rounded corners=5pt] (425pt, 130pt) rectangle (390pt, 100pt) {};
\draw (407pt,115pt) node   {$\mathscr{D}_{q,K}(\cdot)$};

\draw [-latex]    (425pt,115pt) -- (445pt,115pt) ;
\end{scope}

\end{tikzpicture}}
\caption{System model for analog function computation over a noisy MAC. Transmitter $k$ encodes its source block $\mathbf{s}_k$ using the shift-map encoder $\mathscr{E}_q(\cdot)$ to generate a channel input $x_k$. The transmitted signals are superimposed over the channel and corrupted by additive Gaussian noise. The receiver applies decoder $\mathscr{D}_{q, K}(\cdot)$ to recover the desired function outputs~$\hat{u}[m]$.}

    \label{fig:Systemmodel}
\end{figure*}

%% file: Figure/Fig_Coding.tex
\begin{figure}[!t]
    \centering
\scalebox{1}{

\tikzset{every picture/.style={line width=0.75pt}}     

\begin{tikzpicture}[x=0.75pt,y=0.75pt,yscale=-1,xscale=1]

\draw[color=gray, fill=antiquebrass!2 ] (248pt, 90pt) rectangle (226pt, 70pt) {};
\draw[color=gray, fill=antiquebrass!2 ] (226pt, 90pt) rectangle (204pt, 70pt) {};

\draw[color=gray, fill=antiquebrass!15 ] (204pt, 90pt) rectangle (182pt, 70pt) {};
\draw[color=gray, fill=antiquebrass!35 ] (182pt, 90pt) rectangle (160pt, 70pt) {};
\draw[color=gray, fill=antiquebrass!20 ] (138pt, 90pt) rectangle (116pt, 70pt) {};
\draw[color=gray, fill=antiquebrass!40 ] (116pt, 90pt) rectangle (94pt, 70pt) {};

\draw[color=gray, fill=antiquebrass!2 ] (94pt, 90pt) rectangle (72pt, 70pt) {};
\draw[color=gray, fill=antiquebrass!2 ] (72pt, 90pt) rectangle (50pt, 70pt) {};

\draw (237pt,80pt) node {\small$0$};
\draw (215pt,80pt) node {\small $0$};
\draw (193pt,80pt) node {\small $r_{B}$};
\draw (171pt,80pt) node {\small $r_{B-1}$};
\draw (149pt,80pt) node {$\cdots$};
\draw (127pt,80pt) node {\small$r_2$};
\draw (105pt,80pt) node { \small$r_1$};
\draw (83pt,80pt) node {\small $0$};
\draw (61pt,80pt) node {\small $0$};

\draw (72pt,60pt) node {$\overbrace{~~\quad \qquad }^{\beta_m}$};

\draw (227pt,60pt) node {$\overbrace{~~~\quad \qquad }^{\alpha_m-\beta_m}$};

\draw (149pt,60pt) node {$\overbrace{~~\quad\qquad\qquad \qquad\qquad }^{B}$};

\draw (149pt,100pt) node {$\underbrace{\qquad   \qquad \qquad \qquad \qquad \qquad \qquad \qquad \qquad }_{B+\alpha_m}$};

\draw (20pt,80pt) node {$\sum_kt_k[m]:=$};

\draw [-latex]    (130pt,40pt) -- (130pt,55pt) ;
\draw (115pt,45pt) node {$\mathscr{E}(\cdot)$};

\begin{scope}[shift={(4cm,-2cm)}]
    \draw (-60pt,80pt) node {$\sum_k s_k[m]\in $};
    \draw (-30pt,80pt) node {\LARGE$\{$};
\draw[color=gray!30!black!70, fill=amber!10 ] (-15pt,80pt) circle (11pt);
\draw (-15pt,80pt) node { \small $r_1$};

\draw[color=gray!30!black!70, fill=amber!4 ]  (10pt,80pt) circle (11pt);
\draw (10pt,80pt) node { \small $r_2$};

\draw (35pt,80pt) node {$\cdots$};

\draw[color=gray!30!black!70, fill=amber!7 ] (60pt,80pt) circle (11pt);
\draw (60pt,80pt) node { \small $r_{B-1}$};

\draw[color=gray!30!black!70, fill=amber!3 ] (85pt,80pt) circle (11pt);
\draw (85pt,80pt) node { \small $r_{B}$};
\draw (100pt,80pt) node {\LARGE$\}$};
\end{scope}

\end{tikzpicture}}
\caption{Schematic illustration of the proposed shielding mechanism.  The superposition of source symbols is mapped by the encoder $\mathscr{E}(\cdot)$ into a single digit $t_k[m]$ using an enlarged base $B+\alpha_m$, where $\beta_m$ introduces guard intervals and the effective information-bearing region has size $B$. 
This structure prevents carry-over and limits noise propagation across significance levels in the superimposed constellation.}

    \label{fig:Coding}
\end{figure}
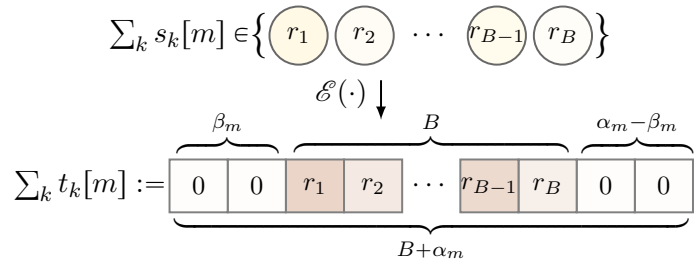